\documentstyle[prl,aps,multicol,psfig]{revtex}
\begin{document}
\draft
\title{Comment on: Weak Anisotropy and Disorder Dependence of the In-Plane Magnetoresistance in
High-Mobility (100) Si Inversion Layers} \maketitle

\begin{multicols}{2}

In a recent paper Pudalov et al. \cite{pud} measured the in-plane
magnetoresistance (MR) of high-mobility (100) Si-inversion layers
and demonstrated that the magnetic field for MR saturation depends
strongly on sample quality. The saturation field found in
experiment "varies for different samples by a factor of 2 at a
given carrier density." They concluded that "the saturation of the
MR cannot be identified with the complete spin polarization of
free carriers". We, however, describe in this comment the
saturation of the MR in the data of Pudalov et al. \cite{pud} by
complete spin polarization of non-localized carriers and the
existence of local moments.

For the interpretation of measurements of the MR for strong
disorder one must take into account the existence of localized
states in the metallic phase. In Ref. \cite{pud} it was argued
that the carrier density $n$ measured by Shubnikov-de Haas
oscillations $n_{SdH}$ and the one measured by the Hall voltage
$n_H$ are nearly identical. But this does not mean that localized
states are not present in metallic samples. The conclusion one can
draw from the relation $n_{SdH}\approx n_H$ is that one cannot see
localized states in the Hall resistivity: apparently the Hall
resistivity is non-critical at the metal-insulator transition,
which occurs at $n=n_c$ or at the  Fermi energy $E_F=E_c$.

In the absence of a magnetic field extended states in a metal have
a spin-degeneracy of two for spin-up and spin-down. Due to Coulomb
interaction, localized states can be singly occupied or doubly
occupied \cite{mott}. At energies close to $E_c$ the density of
singly occupied localized states $n_{so}$ will be small, far from
$E_c$ all localized states are singly occupied. In a two-fluid
model for localized and extended electrons, as already used for
the strongly disordered three-dimensional electron gas in
Phosphorous-doped Silicon \cite{paal}, we assume Curie
paramagnetism for the localized singly occupied states of density
$n_{so}$ and Pauli paramagnetism for the extended states of
density $n-n_{so}$. The singly occupied localized states are
treated as classical spins and are spin-polarized if the
temperature $T$ is sufficiently small: $g^* \mu B>k_B T$. The
other electrons become completely spin-polarized for $B>B_c$ with
$g^* \mu B_c \leq (n-n_{so})/\rho_F$, where $\rho_F$ is the
density of states of the spin-polarized free electron gas given in
terms of the effective mass $m^*$. We note that $B_c \leq
B_{c0}=2E_F/(g^* \mu)$ and due to the existence of localized
states one finds $B_c=0$ for $n=n_{so}$. By taking into account
the modification of screening effects and the density of states
due to spin polarization the MR of the two-dimensional electron
gas for weak disorder was calculated in Ref. \cite{dol}. The
theory doesn't contain localized states and the saturation field
for the MR is $B_{c0}$. For strong disorder \cite{gold} the
saturation field is $B_c$, as calculated above, and the resistance
ratio $R(B>B_c)/R(B=0)$ is strongly increased compared to weak
disorder.

In the experiments of Ref.\cite{pud} the values for $B_c=0$ are
extrapolated from finite $B$values. We find that the slope
$dB_c/dn=dB_{c0}/dn$ does not depend on $n_{so}$, in agreement
with experimental results \cite{pud}, where $dB_c/dn=5.7 T/10^{11}
cm^{-2}$ is found. According to Ref. \cite{mott} one expects that
$n_{so}\approx n_c$ for Silicon MOSFET's with low peak mobility
and $n_{so}\ll n_c$ for more ideal samples, in agreement with
experimental results \cite{ok}. For one sample with very high
mobility Pudalov et al. \cite{pud} also found a larger saturation
field than predicted for an ideal electron gas without localized
states. We don't believe that this is a real effect; the small
shift might be related to the definition of the saturation field
as used to analyze the experimental data or related to the
relatively high temperature of the experiment.

In conclusion we argue that the MR-experiments of Pudalov et al.
\cite{pud} give indications for local moment formation in the
two-dimensional electron gas in the metallic phase.

V.T. Dolgopolov acknowledges support of RFBR via grants
00-02-17294, 01-02-16424, and  by  program "Nanostructures" from
the Russian Ministry of Sciences.

V.~T. Dolgopolov $^1$ and A.Gold $^2$

$^1$ {\it Institute of Solid State Physics, Chernogolovka, Moscow
District 142432, Russia}

$^2$ {\it Centre d`Elaboration de Mat\'eriaux et d`Etudes
Structuales (CEMES-CNRS), 29 Rue Jeanne Marvig, 31055, Toulouse,
France}


\end{multicols}

\end{document}